\documentclass{PoS}

\newcommand{\psibar}{\bar{\psi}}
\newcommand{\Udag}{U^{\dagger}}

\newcommand{\gi}{\gamma_i}

\newcommand{\ihat}{\hat{\imath}}

\usepackage{amsmath, graphicx}
\usepackage{caption,multicol}
\usepackage{float}

\title{Spectral functions of charmonium from 2 flavour anisotropic lattice data}
\ShortTitle{Charmonium spectral functions}
\author{\speaker{Aoife Kelly}\\
        Department of Mathematical Physics, 
        National University of Ireland Maynooth,
        Maynooth,
        Co. Kildare,
        Ireland.\\
        E-mail: \email{aoifek@thphys.nuim.ie}}

\author{Jon-Ivar Skullerud\\
        Department of Mathematical Physics,
        National University of Ireland Maynooth,
        Maynooth,
        Co. Kildare,
        Ireland.\\
        E-mail: \email{jonivar@thphys.nuim.ie}}

\author{Chris Allton\\
        Department of Physics,
        Swansea University,
        Swansea, SA2 8PP
        Wales,
        UK\\        
        E-mail: \email{c.allton@swan.ac.uk}}

\author{Dhagash Mehta\\
        Department of Mathematics,
        North Carolina State University,
        Raleigh,
        NC 27695-8205,
        USA\\
        E-mail: \email{dbmehta@ncsu.edu}}

\author{Mehmet B. Oktay\\
        Department of Physics and Astronomy,
        University of Iowa,
        Iowa City,
        Iowa 52242
        USA\\
        Email: \email{boktay@ucair.med.utah.edu}}

\abstract{
The spectral functions of QCD can give us insight into properties of hadrons, and they are useful in probing the QCD vacuum. I will discuss the correlators and spectral functions of charmonium in high temperature two flavour QCD. The spectral functions have been obtained using the Maximum Entropy Method from anisotropic lattice data using the conserved vector current. This work has been done as part of the FASTSUM collaboration. We find that the spectral functions for zero momentum are stable. At non-zero momentum the spectral functions are less stable but still produce resonance and transport peaks. This work is part of our programme to calculate the heavy quark diffusion constant.
}

\FullConference{31st International Symposium on Lattice field theory LATTICE 2013\\
         July 29 August 3, 2013\\
         Mainz, Germany}

\begin{document}

\section{Introduction}
Experimental observations have been made of the quark-gluon plasma at both RHIC and the LHC. As yet, the theory explaining this phase of matter has not been fully developed. The charm quarks were seen to thermalise as effectively as the light quarks. This motivates the study of charm diffusion \cite{Petreczky:2006} and $J/\psi$ suppression \cite{Matsui:1986}. There have been several previous lattice studies of charmonium at high temperature \cite{Aarts:2007, Asakawa:2003}. Hydrodynamical transport coefficients are related to the low frequency range of the spectral function by Kubo formulae \cite{Forster:1990, Boon:1980}. The heavy quark diffusion can be related to the behaviour of vector current correlators at large times. In order to calculate this diffusion, the correlator is decomposed into its transverse and longitudinal components \cite{Petreczky:2006}. The heavy quark diffusion constant $D$ is related to the spatial components of the vector spectral function,
\begin{equation}\label{eq:diffusion}
D = \frac{1}{6\chi^{00}}\lim_{\omega\rightarrow 0}\sum_{i=1}^3\frac{\rho_{ii}^{V}(\omega,\vec{p}=0,T)}{\omega},
\end{equation}
where $\chi^{00}$ is the quark number susceptibility which is defined through the zeroth component of the temporal correlator in the vector channel \cite{Ding:2012}. The density-density retarded correlator $\chi_{NN}(\vec{p},\omega)$ is also related to the heavy quark diffusion constant \cite{Petreczky:2006}. 
\begin{equation}\label{eq:diffcorr}
\chi_{NN}(\omega,\vec{p}) = \frac{\chi_s(\vec{p})Dp^2}{-i\omega+p^2D}-\frac{\chi_s(\vec{p})Dp^2}{-i\omega+\eta}
\end{equation}
Here $\chi_s(\vec{p})$ is the static susceptibility and $\eta$ is the drag coefficient. Additional information may thus be obtained by studying the spectral functions of charmonium for non-zero momentum. For frequencies $\omega \sim Dp^2$ the first term dominates and this represents the solution to the diffusion equation.  

Different methods have been used to compute the spectral functions of charmonium. One such method includes the use of the Maximum Entropy Method (MEM) \cite{Asakawa:2001} and phenomenologically motivated fit functions \cite{Ding:2012}. Our approach is to use MEM analysis to obtain the spectral functions for charmonium in the vector channel, $J/\psi$ and to compute the heavy quark diffusion constant. We use the updated Bryan's method for this \cite{Aarts-Allton:2007}. This work has been performed with anisotropic lattices with two dynamical flavours of fermions. 

\section{Lattice Parameters}
The lattices used were generated \cite{Oktay:2010} with two dynamical flavours of light quarks with aniso\-tropy $\xi = a_s/a_{\tau} = 6$. The spatial and temporal lattice spacings are $a_s \simeq 0.162$fm and $a_{\tau}^{-1} \simeq 7.35$GeV. The temperature, $T = 1/(a_{\tau}N_{\tau})$, ranges from the confined phase up to $\sim 2T_c$, where $T_c$ is the deconfining transition. The fermion action is a Hamber-Wu action in the spatial directions and a Wilson action in the temporal direction. A two-plaquette Symanzik-improved gauge action was employed. The lattices, their corresponding temperatures and the number of configurations are listed in Table \ref{tab:lattice}.

\begin{table}[h!]
\begin{center}
\begin{tabular}{ccccc}
\hline
$N_s$ & $N_{\tau}$ & $T$ (MeV) & $T$/$T_c$ & $N_{\text{cfgs}}$ \\
\hline \hline
12 & 16 & 459 & 2.09 & 1000 \\
12 & 18 & 408 & 1.86 & 700 \\
12 & 20 & 368 & 1.68 & 1000 \\
12 & 24 & 306 & 1.40 & 500 \\
12 & 28 & 263 & 1.20 & 1000 \\
12 & 32 & 230 & 1.05 & 875 \\
\hline
\end{tabular}
\caption{Lattice sizes $N_s^3\times N_{\tau}$, temperatures $T$ and number of configurations $N_{\text{cfgs}}$.}
\label{tab:lattice}
\end{center}
\end{table}

\section{Spectral Functions from MEM}
The Maximum Entropy Method (MEM) was implemented in order to find the spectral functions for the $J/\psi$ particle. This is a well-established method \cite{Asakawa:2001} of determining the spectral function $\rho(\omega)$ from the imaginary time correlator $G(\tau)$,

\begin{equation}\label{eq:correlator}
G(\tau) = \int\limits_0^{\infty}\frac{d\omega}{2\pi}\rho(\omega)\frac{\cosh[\omega(\tau-\frac{\beta}{2})]}{\sinh(\frac{\beta\omega}{2})},
\end{equation}

\noindent for $\beta = 1/T$. MEM is contingent on Bayes' Theorem and produces the most probable spectral function given the Monte Carlo data.

The resulting spectral functions rely on certain input information known as the default models $m(\omega)$. The inclusion of this prior information leads to a unique solution. However, while it is necessary to include some prior information in these $m(\omega)$, the ensuing spectral function should have no dependence on them. We have studied the default models listed in equations \eqref{eq:dm0}--\eqref{eq:dm5}, where $m_0$ is a free parameter and $m_1$ is scaled with temperature,

\begin{align}
m(\omega) = m_0\omega^2, \label{eq:dm0} \\[0.3cm]
m(\omega) = m_0, \label{eq:dm1} \\[0.3cm]
m(\omega) = m_0\omega(m_1 + \omega). \label{eq:dm5}
\end{align}

The parameter $m_0$ was chosen to be the same for each temperature $T$, close to the value giving a best fit to the correlator. It was then varied by factors of 10 and 1/10 in order to examine the dependence of $m(\omega)$ on it. The parameter $m_1$ was set to $1/T$ and then varied to $2\pi T$ and $T/2\pi$. The analysis in the variation of $m_1$ is ongoing. Most of the results shown are using \eqref{eq:dm5}.

\section{Conserved Vector Current}
The focus of this study is the vector channel, $J/\psi$. The conserved current was used to create this state and has the advantage that it requires no renormalisation. The spatial component of the conserved current for the Hamber-Wu action is given in equation \eqref{eq:spatialCVC}. Here $r_A = 6s = 6/8$, where $s=1/8$ is the spatial Wilson parameter. The temporal component is the usual conserved Wilson current, which is described in equation \eqref{eq:temporalCVC}, where $r=1$ is the usual Wilson parameter.


\begin{equation}\label{eq:spatialCVC}
\begin{split}
V_i^{A}(x) = &
  -\frac{2}{3}\psibar(x)(r_A-\gi)U_i(x)\psi(x+\ihat)
  +\frac{2}{3}\psibar(x+\ihat)(r_A+\gi)\Udag_i(x)\psi(x) \\
  & +\frac{1}{12u_s}\Big[
   \psibar(x)(2r_A-\gi)U_i(x)U_i(x+\ihat)\psi(x+2\ihat)
  + (x\to x-\ihat)\Big] \\
  & -\frac{1}{12u_s}\Big[
  \psibar(x+\ihat)(2r_A+\gi)\Udag_i(x)\Udag_i(x-\ihat)\psi(x-\ihat)
  + (x\to x+\ihat)\Big]\,.
\end{split}
\end{equation}

\begin{equation}\label{eq:temporalCVC}
V_t(x) = \frac{1}{2}[\bar{\psi}(x+\hat{t})(r+\gamma_0)U_t^\dagger(x)\psi(x) - \bar{\psi}(x)(r-\gamma_0)U_t(x)\psi(x+\hat{t})].
\end{equation}

In order to calculate the heavy quark diffusion constant \cite{Petreczky:2006}, it is necessary to decompose the conserved vector current into its longitudinal ($V_L$) and transverse ($V_T$) polarisations.
\begin{equation}\label{eq:polarisation}
V_{ij}(\tau,\vec{p}) = \left(\delta_{ij}-\frac{p_ip_j}{p^2}\right)V_T(\tau,\vec{p}) + \frac{p_ip_j}{p^2}V_L(\tau,\vec{p}).
\end{equation}
It is the longitudinal component which will yield the charm diffusion. The momentum values here are given by $p^2 = \left(\frac{2\pi n^2}{a_sN_s}\right)^2$, with $n^2$ = 0, 1, 2, 3, 4, corresponding to $p$ = 0, 0.66, 0.93, 1.14, 1.32 GeV respectively.

\section{Zero Momentum Results}
\begin{figure}[H]
\begin{center}
\includegraphics[width=0.62\textwidth]{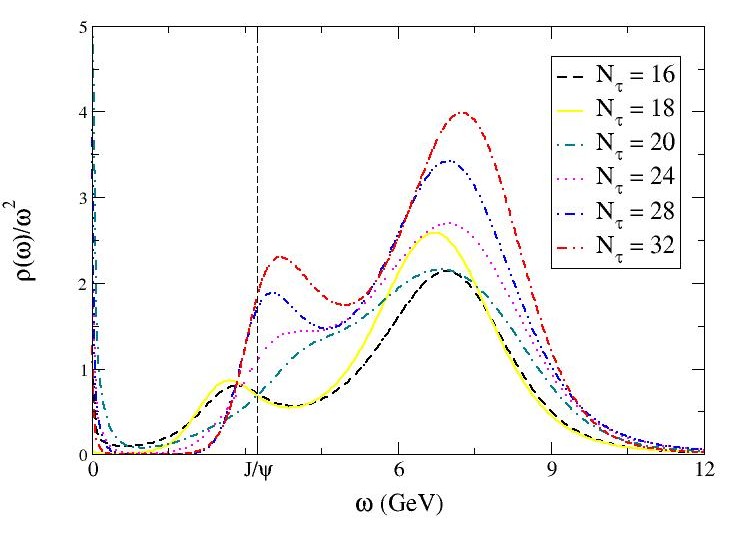}
\caption{Spectral functions for each temperature with $p^2$ = 0. The vertical line represents the experimental value for the $J/\psi$ mass.}
\label{fig:zeromom}
\end{center}
\end{figure}

Figure \ref{fig:zeromom} shows the spectral functions obtained for momentum $p^2 = 0$ for each temperature. The default model used in these calculations is given in equation \eqref{eq:dm5}. A peak structure is evident (in particular for $N_{\tau}$ = 28, 32) close to the experimental value of the $J/\psi$ mass. For the higher temperatures, we see no evidence of a peak. The secondary peak can be attributed to lattice artifacts.

\section{Non-zero Momentum Results}
\begin{figure}[H]
\begin{center}
\includegraphics[width=0.5\textwidth]{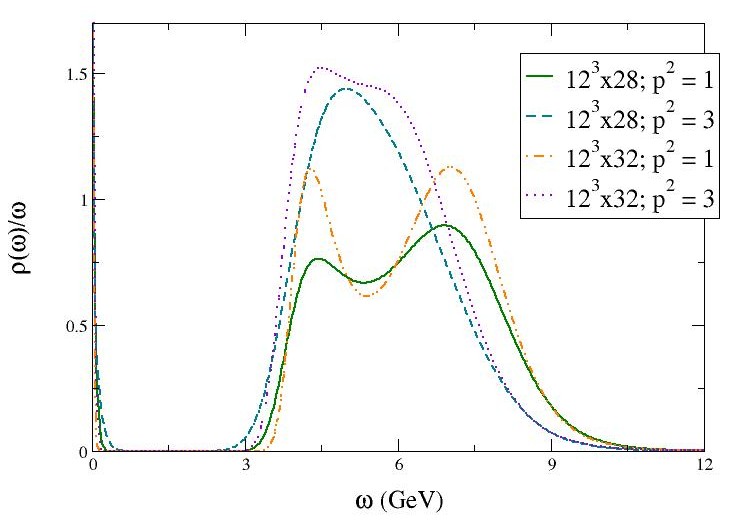}
\captionof{figure}{Spectral functions for the longitudinal polarisation of the conserved vector current at non-zero momenta.}
\label{fig:longmomenta}
\end{center}
\end{figure}

In studying the non-zero momenta it was noted that the high temperature lattices did not yield good results. The systematic errors were quite large. This has been already observed in \cite{Oktay:2010}. As such, only the $12^3\times28$ and $12^3\times32$ lattices are shown here. Figure \ref{fig:longmomenta} shows the spectral functions obtained for the longitudinal polarisation of momenta $p^2 = 1$ and $p^2 = 3$. At $p^2 = 1$ there is a peak structure detectable, which is non-existent at $p^2 = 3$. However, we cannot definitively argue that this is due to a melted state. Further analysis is required.
\begin{multicols}{2}
\columnbreak

\begin{figure}[H]
\begin{center}
\includegraphics[width=0.5\textwidth]{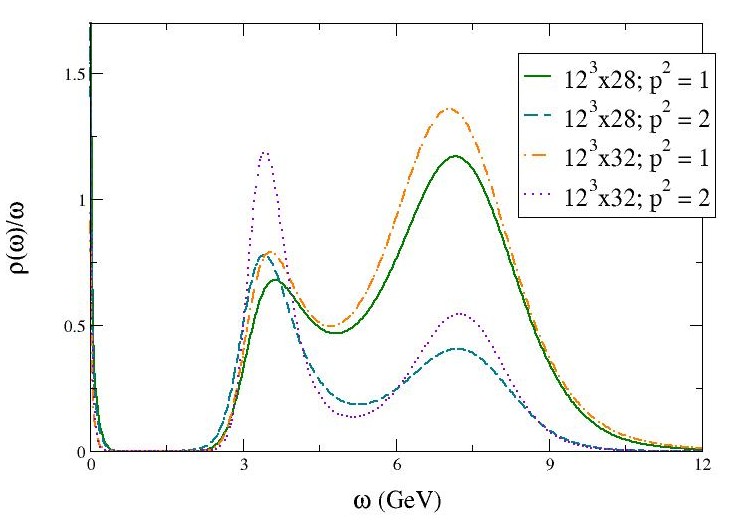}
\captionof{figure}{Spectral functions for the transverse polarisation of the conserved vector current at non-zero momenta.}
\label{fig:tranmomenta}
\end{center}
\end{figure}
\end{multicols}

In figure \ref{fig:tranmomenta} we present the plots for the spectral functions of the transverse component of the conserved vector current for momenta $p^2 = 1$ and $p^2 = 2$. At these low momenta we see evidence of a stable peak in the region of the $J/\psi$ mass, with a secondary peak due to lattice artifacts. 

\begin{figure}[h]
\begin{center}
\includegraphics[width=0.6\textwidth]{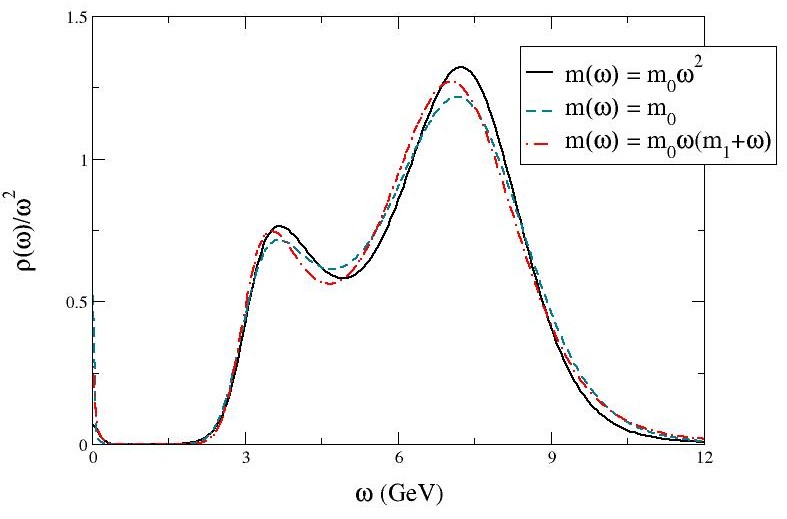}
\captionof{figure}{Dependence of $\rho(\omega)$ on $m(\omega)$ for $p^2 = 0$ on the $12^3\times32$ lattice. }
\label{fig:dmdependence}
\end{center}
\end{figure}

When using MEM it is prudent to check the dependence of the spectral function on the default model. Thus, in our studies of the charmonium spectral functions three default models were used, see equations \eqref{eq:dm0} -- \eqref{eq:dm5}. The dependence is shown in figure \ref{fig:dmdependence} on the $12^3\times28$ lattice for the transverse polarisation of the conserved current with $p^2 = 1$. As the spectral functions have similar shapes with peak positions very close to one another, we can be reasonably confident in our results. 

\section{Low Frequency Region}
\begin{figure}[h]
\begin{center}
\includegraphics[width=0.6\textwidth]{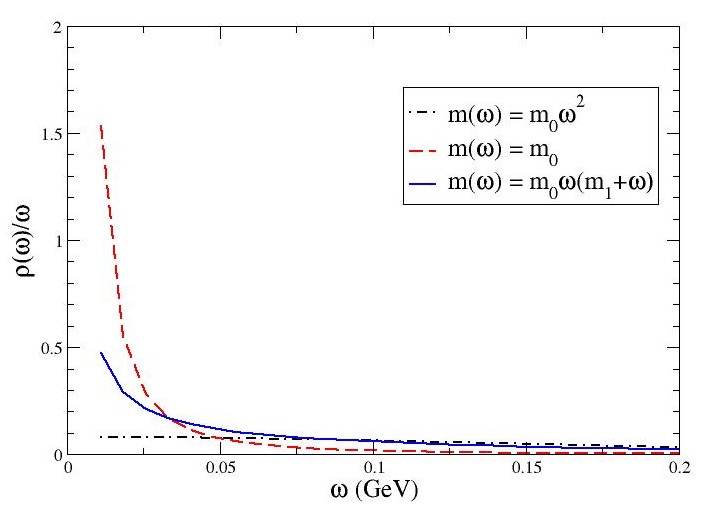}
\caption{Low frequency region of spectral function from $12^s\times32$ lattice with zero momentum.}
\label{fig:lowfreq}
\end{center}
\end{figure}

Our objective is to calculate the diffusion transport coefficient and therefore we are particularly interested in the low frequency region of our spectral functions. Figure \ref{fig:lowfreq} shows the low frequency zone for the spectral function the $12^3\times32$ lattice with $p^2 = 0$ for each default model. There are discrepancies between the heights of these transport peaks, however the range is small. It is these peaks which we will use to find the coefficient of charm diffusion. As $\omega \rightarrow 0$ the default model in equation \eqref{eq:dm0} (represented on the plot by the black line) will converge to 0, and the default model in equation \eqref{eq:dm1} (represented by the red line) will diverge. These therefore will give extreme upper and lower bounds. Our main result is given by the blue line. A variation in values for $m_0$ and $m_1$ has also been studied, and the dependence of the spectral functions on these parameters is in progress.

\section{Conclusions and Outlook}
We have computed the spectral functions in the vector channel (corresponding to $J/\psi$) from lattice QCD using anisotropic lattices with the conserved vector current. At zero momentum, the calculations agree quite well and confirm previous studies suggesting that $J/\psi$ survives up to at least $1.2T_c$. At non-zero momentum, the errors are larger and we are less confident with our results. 

We are particularly interested in the low frequency region of the spectral functions. In our future work the transport coefficients will be evaluated, including the heavy quark diffusion and the susceptibility. Lattices with 2+1 dynamical flavours of quarks and a finer spatial lattice spacing will also be examined.

\section*{Acknowledgements}
This work was supported by the Irish Research Council and Science Foundation Ireland grant 08-RFP-PHY1462. We acknowledge the use of computing resources from the Irish Centre for High End Computing, Fermilab Lattice QCD Clusters, the Trinity Centre for High Performance Computing and the IITAC project funded by the HEA under the Program for Research in Third Level Institutes (PRTLI) co-funded by the Irish Government and the European Union. DM was supported by a DARPA Young Faculty Award and the U.S. Department of Energy under contract no. DE-FG02-85ER40237.


\begin{thebibliography}{99}
\bibitem{Petreczky:2006}
  P. Petreczky and D. Teaney,
  \emph{Heavy quark diffusion from the lattice},
  Phys. Rev. D \textbf{73} (2006) 014508 [arXiv:0507318 [hep-lat]].

\bibitem{Matsui:1986}
  T. Matsui and H. Satz,
  \emph{$J/\psi$ suppression by quark-gluon plasma formation},
  Phys. Lett. B \textbf{178} (1986) 416.

\bibitem{Aarts:2007}
  G. Aarts, C. Allton, M. B. Oktay, M. Peardon and J.-I. Skullerud,
  \emph{Charmonium at high temperature in 2 flavour QCD},
  Phys. Rev. D \textbf{76} (2007) 094513 [arXiv:0705.2198 [hep-lat]].

\bibitem{Asakawa:2003}
  M. Asakawa and T. Hatsuda,
  \emph{Charmonia above the deconfinement phase transition}
  [arXiv:0309001 [hep-lat]]

\bibitem{Forster:1990}
  D. Forster,
  \emph{Hydrodynamics, fluctuations, broken symmetry and correlation functions},
  Perseus Books (1990).

\bibitem{Boon:1980}
  J.P. Boon and S. Yip,
  \emph{Molecular Hydrodynamics},
  McGraw-Hill (1980).

\bibitem{Ding:2012}
  H.-T. Ding, A. Francis, O. Kaczmarek, F. Karsch, H. Satz and W. Soeldner,
  \emph{Charmonium properties in hot quenched lattice QCD},
  Phys. Rev. D \textbf{86} (2012) 014509.

\bibitem{Asakawa:2001}
  M. Asakawa, Y. Nakahara and T. Hatsuda, 
  \emph{Maximum entropy analysis of the spectral functions in lattice QCD},
  Prog. Part. Nucl. Phys. \textbf{46} (2001) 459 [arXiv:0011040 [hep-lat]].


\bibitem{Aarts-Allton:2007}
  G. Aarts, C. Allton, J. Foley and S. Hands,
  \emph{Spectral functions at small energies and the electrical conductivity in hot, quenched lattice QCD},
  Phys. Rev. Lett. \textbf{99} (2007) 022002 [arXiv:0703008 [hep-lat]]

\bibitem{Oktay:2010}
  M. B. Oktay and J.-I. Skullerud,
  \emph{Momentum dependence of charmonium spectral functions from lattice QCD},
  [arXiv:1005.1209 [hep-lat]].


\end{thebibliography}
\end{document}